\input amstex.tex
\input amsppt.sty
\magnification1200
\documentstyle{amsppt}
\NoBlackBoxes
\hcorrection{20 true mm}
\vcorrection{30 true mm}
\pagewidth{30 true pc}
\pageheight{47 true pc}

\document
\loadeurm
\loadbold

\topmatter
\title
The new interpretation of~quantum mechanics and hidden parameters
\footnotemark[1,]
\footnotemark[2]
\endtitle

\rightheadtext{New interpretation of~quantum mechanics}

\author
Ji\v{r}\'{\i} Sou\v{c}ek\\
\endauthor

\affil
\it Charles University, Prague
\endaffil

\address
\newline
Faculty of Mathematics and Physics
\newline
Sokolovsk\'a 83
\newline
186 00  Praha 8
\newline
Czech Republic
\medskip
\endaddress

\email
soucekj\@karlin.mff.cuni.cz
\endemail

\footnotetext[1]{A part of~this paper was presented at~the~Conference "Hadron Structure
'76" held in~Smole\-ni\-ce (Czechoslovakia) 15.-19.11.1976. A~revised form
can be found in~[14].}
\footnotetext[2]{This paper is an~identical copy of~preprint No. NBI-HE-81-4
of~Niels Bohr Institute (Copenhagen) from November 1981. It has never
been published regularly.}

\thanks
Preparation of this paper was supported by the~Grant No. RN\,19982003014
of~the~Ministry of~Education.
\endthanks

\abstract
The new interpretation of~quantum mechanics is based on~a~complex
probability theory. An~interpretation postulate
speci\-fies events which
can be observed and~it follows that the~complex probability of~such
event is, in fact, a~real positive number. The~two-slit experiment,
the~mathematical formulation of~the~complex probability theory,
the~density matrix, Born's law and~a~possibility of~hidden variables are
discussed.
\endabstract

\endtopmatter


\head
1. Introduction
\endhead

In this paper the old problem of~Quantum Mechanics (QM) --
the~interference of~probability amplitudes -- is investigated from
the point of~view of~a~complex theory of~probability (C-TP). It can be
said that the~"mystery of~QM" is not explained until now, if we
understand under the~term "to~explain" something deeper than "to say how
to~calculate observed numbers".

Here we suggest a~way to~imagine what is going to~happen in~the two-slit
experiment; the~calculated numbers, however, are the~same as before.

QM is necessarily a~probabilistic theory, but the~probability is used
here in a~completely non-classical way [1]. The~role of~probability is
played by~the~probability amplitude, but this amplitude is complex
whereas the probability must be real.

It is well recognized that it is virtually impossible to~ascribe
a~de\-finite trajectory to~the~electron, but it is equally impossible
to~ascribe a~set of~trajectories having different probabilities to~it;
in~this respect the~electron differs from a~Brownian particle. It is
possible to~assign the~Feynman probability amplitude to~each trajectory,
to~add them, and at~an~appropriate moment to~take the~squared modulus
using the~RPI (rule of~principal indistinguishability): for~principally
indistinguishable alternatives add the~amplitude, for distinguishable
ones add the~probabilities [1], [2].

For an electron we suggest the~following concept of~the~trajectory: it
is a~couple $(\gamma_+, \gamma_-) = \gamma$ of~two possible
trajectories, where $\gamma_+$ and~$\gamma_-$ are oriented forward
and~backward in~time, respectively. To~each such "trajectory" there
corresponds a~complex probability $\Phi$ given by
$$
\Phi (\gamma) = \phi (\gamma_+) [\phi (\gamma_-)]^*, \tag 1
$$
where $\phi (\gamma_+)$ is the~Feynman amplitude for the~path
$\gamma_+$.
Equation (1) in~a~way resembles the~defionition of~a~density matrix
$\rho (x', x) = \psi (x') \psi^* (x)$. It can be said that
the~electron moves independently in~both directions of~time with
generally different trajectories.

Now, the two-slit experiment (Exp.\,1) can be simply explained. There
are four possible ways to~pass from $s$ to~$x$ through slits 1 and~2:
\roster
\item"(i)" $\gamma_+$ and $\gamma_-$ go through 1,
\item"(ii)" $\gamma_+$ and $\gamma_-$ go through 2,
\item"(iii)" $\gamma_+$ goes through 1 and $\gamma_-$ goes through 2,
\item"(iv)" $\gamma_+$ goes through 2 and $\gamma_-$ goes through 1.
\endroster
Diagramatically it is written as (Fig.\,1).

Note that this means decomposition of~the~event
$(s \rightarrow x)_{through\ 1+2}$ into four disjunctive subevents (!);
really no numbers enter Fig.\,1.

Let us suppose that the~observation of~an~electron at~a~point $P$ means
that both $\gamma_+$ and $\gamma_-$ pass through $P$ (so called {\it
interpretation postulate}). Let us consider the~second experiment (Exp.\,2)
in~which the~electron is observed at~slit 2. From
the~interpretation postulate it follows that the~possibilities (iii)
and~(iv) are excluded and~we have (Fig.\,2).
Thus the C-probability of~an~event $(s \rightarrow x)_{through\ 1+2}$
is given by
$$
\multline
\Phi_{Exp.\,1} (x) = \Big( \sum_{\gamma \in (\roman i)} + \dots +
\sum_{\gamma \in (\roman{iv})} \Big) \Phi (\gamma) = \\
= | \langle x | s \rangle_1
|^2 + | \langle x | s \rangle_2 |^2 + 2 Re \langle x | s\rangle_1
\langle x | s \rangle_2^*
\endmultline
$$
$$
\Phi_{Exp.\,2} (x) = \Big( \sum_{\gamma \in (\roman i)} + \sum_{\gamma \in
(\roman{ii})} \Big) \Phi (\gamma) = | \langle x | s \rangle_1 |^2 +
| \langle x | s \rangle_2 |^2,
$$
because, for example
$$
\sum_{\gamma \in (\roman{iii})} = \Big( \sum_{\gamma_+ \ni 1} \phi (\gamma_+)
\Big) \Big( \sum_{\gamma_- \ni 2} \phi (\gamma_-) \Big)^* =
\langle x | s \rangle_1 \langle x | s \rangle_2^*.
$$

The cases (iii) and (iv) may be called {\it the~interference events}. In
our theory the~difference between Exp.\,1 and Exp.\,2 does not arise
because of~a~disturbance of~the~electron by~the~measuring aparatus, but
in~our two experiments we observe truly different events.

The interpretation postulate thus replaces RPI; IP is a~natural element
of~probabilistic description of~Nature: this is the~rule determining
which "event" in~the~theory corresponds to~the~event observed in~an
experiment.

The correspondence between the~classical particle going through
the~point $x$ and its mathematical description by~a~curve containing $x$
is considered as self-evident. The~correspondence between an~electron
passing through $x$ and~its description by~$\gamma = (\gamma_+,
\gamma_-)$ with $x \in \gamma_+ \cap \gamma_-$ may be considered
as strange, but we think this strangeness is apparent, induced by~our
classical background.

The~concept of~$\gamma = (\gamma_+, \gamma_-)$ contains a~certain
element of~non-locality; this is exactly what is observed in~QM -- see
the~Einstein-Rosen-Podolski paradox [3]. For example, the~question "how
can the~electron passing through slit~1 know whether the~other slit
is closed or open?" can be answered simply: the~electron may find out
this using its backward trajectory $\gamma_-$ (supposing $\gamma_+$
passes through slit 1). If both slits are open, the~electron can use
possibilities (iii) and~(iv); these possibilities cannot apply if one or
other of~the~slits is closed. The~interference character (i.e. so-called
wave properties) of~electrons are natural in~C-TP; the~usual R-TP is
rather of~an~"adding" character.
$$
$$

\head
2. Complex theory of probability
\endhead

Brownian motion can be mathematically described [4] by~the~R-TP
$(\Omega, \Sigma, P)$, where
$$
\align
\Omega &= \text{ space of~elementary events } = \\
&= \{ \text{all possible trajectories of~a~Brownian particle}\} = \\
&= \Big\{ \gamma: \Bbb R \rightarrow
\Bbb R^4 \big|\ \gamma (\tau) = \big[ t (\gamma (\tau)), \vec x
(\gamma (\tau)) \big], \frac{d}{d \tau} t (\gamma (\tau)) > 0 \Big\},\\
\Sigma &= \text{ system of~events -- system of~measurable subsets of~}
\Omega,\\
P &= \text{ non-negative } \sigma \text{-additive measure on } \Sigma,\
P(\Omega) = 1.
\endalign
$$
We shall use the~symbolic notation $P(\gamma)$ (exact for $\Omega$
finite). The probability of~the~transition $[x_1 \rightarrow x_2]$,
$x_1, x_2 \in \Bbb R^4$ is then given by
$$
P [x_1 \rightarrow x_2] = \sum_{\gamma(0) = x_1, \gamma(1) = x_2}
P(\gamma),
$$
where $P(\gamma)$ is the Wiener measure on~$\Sigma$.

Now we shall give the~exact definition of~the~notions introduced
in~the~Introduction.

\definition{Definition}
The {\it trajectory} of~quantum particle is a~couple
$\gamma = \mathbreak
(\gamma_+, \gamma_-)$ of~two oriented curves in~space-time, $\gamma_+,
\gamma_-: \Bbb R \rightarrow \Bbb R^4$, such that $\frac{d}{d \tau}
t(\gamma_+(\tau)) > 0$, $\frac{d}{d \tau} t(\gamma_-(\tau)) < 0$. (In this
paper we consider the~case of~non-relativistic QM.) {\it C-TP} is
the~system $(\Omega, \Sigma, \Phi)$, where $\Omega = \{ \gamma =
(\gamma_+, \gamma_-) \}$, $\Sigma$ = system of~"measurable" subsets
of~$\Omega$ and~$\Phi$ is a~complex
function defined on~$\Omega$ satisfying axioms (A1)--(A4) given below.

The {\it trajectory $\gamma^+$ adjoint to~$\gamma = (\gamma_+, \gamma_-)$}
is defined by $\gamma^+ = (\check\gamma_-, \check\gamma_+)$, where
$\check\gamma_{\pm}(\tau) = \gamma \pm (- \tau)$;
$E^+ = \{\gamma^+ |\ \gamma \in E \}$ for $E \in \Sigma$ (and $E^+ \in
\Sigma$ is supposed).

The event $E \in \Sigma$ is called {\it hermitian} ($E \in\Sigma_{herm}$)
if~$E^+ = E$.

Let us denote $\Omega_+ = \{ \gamma_+ |\  (\gamma_+, \gamma_-) \in \Omega
\}$, $\Omega_- = \{ \gamma_- \}$. The event $E \in \Sigma$ is called
{\it pure} ($E \in\Sigma_{pure}$), if $E = A \times \check A$, where $A
\subset \Omega_+$, $\check A = \{ \check\gamma |\ \gamma \in A \}$ and~$A \times
\check A$ means
the~Cartesian product.

The event $E \in \Sigma$ is called {\it mixed} ($E \in \Sigma_{mix}$)
if~$E$ is a~disjoint union of~pure events.
\enddefinition

$\Phi$ is assumed to~satisfy the~following axioms.

\roster
\item"(A1)"
$\Phi$ is a~$\sigma$-additive complex measure on~$\Sigma$,
i.e., symbolically, $\Phi (E) = \sum_{\gamma \in E} \Phi(\gamma)$, $E
\in \Sigma$,
\item"(A2)"
$\Phi (\gamma^+) = [ \Phi (\gamma) ]^*$, i.e. $\Phi (E^+) = [ \Phi (E) ]^*$,
\item"(A3)"
$\Phi (\gamma_+, \gamma_-) =  \Phi (\gamma_+ \times \Omega_-) \cdot
\Phi (\Omega _+ \times \gamma_-)$ or, more generally,
 $\Phi (A \times B) = \Phi (A \times \Omega_-) \cdot
\Phi (\Omega _+ \times B)$.
\endroster
Thus it suffices to know $\Phi$ for events of~type $\gamma_+ \times
\Omega_-$. This axiom implies the~statistical independence of~$\gamma
_+$ and~$\gamma_-$. Moreover, in~QM the~following rules hold:
\roster
\item"(A4)"
$\Phi(\gamma_+ \times \Omega_-) = \phi (\gamma_+) =
e^{iS(\gamma_+)} = $ Feynman's amplitude for~$\gamma_+$.
\endroster
If~$E = A \times \check A \in \Sigma_{pure}$, using (A3), (A2) and~(A4) we obtain
$$
\spreadlines{-4pt}
\multline
\Phi (E) = \Phi (A \times \Omega_-) \cdot \Phi(\Omega_+ \times \check{A}) =\\
= \Phi (A \times \Omega_-) \cdot \big[ \Phi(A \times \Omega_-) \big]^* =
\Big| \sum_{\gamma_+ \in A} \phi(\gamma_+) \Big|^2.
\endmultline
$$
Thus the C-probability of~pure events is (after appropriate
normalization) equal to~the~usual real probability. The~same is true for
mixed events, because $\Phi(E) = \sum \Phi (E_k)$ if~$E =
\bigcup_{disj} E_k$, $E_k \in \Sigma_{pure}$. From (A3) we have (for $A =
\Omega_+$, $E = \Omega_-$): $\Phi (\Omega) = |\Phi (\Omega)|^2$ and thus
$\Phi (\Omega) = 1$ (assuming $\Phi (\Omega) \not= 0$).
\smallskip

{\it Interpretation postulate} (IP).

Let us suppose that the~experiment is
prepared in such a~way that  the~presence of~the~electron
at~(space-time) points $x_1, \dots, x_n$, $y_1, \dots, y_m$ is measured.
Let us consider an~event in~which the~presence of~an~electron
at~$x_1, \dots, x_n$ was confirmed and~at~$y_1, \dots, y_m$ excluded.
This situation is described in~our theory as an~event
$$
\multline
E_{x, y} = E_{x_1, \dots, x_n,\ y_1, \dots, y_m} := \\
= \big\{ (\gamma_+, \gamma_-) | \ x_1, \dots, x_n \!\in\! \gamma_+ (\Bbb R)
\cap \gamma_- (\Bbb R), \ y_1, \dots, y_m \!\in\! \gamma_+ (\Bbb R) \cup \gamma_-
(\Bbb R) \big\}.
\endmultline
$$

We can assume that all observations and~preparations of~quantum systems
may be described in~terms of~$E_{x, y}$. We see that $E_{x, y}
\in \Sigma_{pure}$ and
$$
\Phi(E_{x, y}) = \big| \!\!\!\!\!\!\!\!\! \sum_{\Sb
\gamma_+ \ni x_1, \dots, x_n,\\
\gamma_+ \not\ni y_1, \dots, y_m
\endSb}
\phi(\gamma_+) \big|^2.
$$

Generally, the observable events are of~the~type $\Sigma_{mix}$ while
the~dynamics is given in~terms of~$\phi(\gamma_+) = \Phi (\gamma_+
\times \Omega_-)$ alone. This is the~reflection of~the~fact that in~QM
the~observed quantities (= probabilities) are expressed bilinearly
in~dynamical quantities (amplitudes or~wave functions).

$$
$$

\head
3. The density matrix
\endhead

Our discussion of~the~two-slit experiment was simplified because
the~experiment was described in~space variables only. Here
the~transition to~the~space-time description $x = (t, \vec x)$, $S_1
= (t_1, \vec x_1)$ etc. will be assumed. Let us suppose that
in~the~$n$-slit experiment the~electron is measured at~slits $S_1, \dots,
S_m$, ($0 \leq m \leq n$). The~event observed is then given by
$$
\align
E(x) &= \Big( \bigcup_{k = 1}^m E_{kk} \Big) \cup
\Big( \bigcup_{k, l = m + 1}^n E_{kl} \Big),\\
E_{kl} &= \big\{ (\gamma_+, \gamma_-) |\  s, x \in \gamma_+ (\Bbb R) \cap
\gamma_- (\Bbb R) ,\ S_k \in \gamma_+ (\Bbb R) ,\ S_l \in \gamma_- (\Bbb R)
\big\}
\endalign
$$
and its probability by
$$
\align
\Phi (E(x)) &= \sum_{k = 1}^m | \phi_k |^2 + \sum_{k, l = m + 1}^n \phi_k
\phi_l^*,\\
\phi_k &= \sum_{x, s, S_k \in \gamma_+(\Bbb R)} \phi (\gamma_+).
\endalign
$$

Let us suppose that the~electron was observed at~$x_0$ at~the~moment
$t_0$, then it has passed through a~slit system with some slits equipped
with measuring apparatus. This is so-called {\it preparation}
of~the~electron. The~corresponding event
$$
E = \bigcup_{k = 1}^n A_k \times \check A_k \in \Sigma_{mix}
$$
is defined as the~set of~all "trajectories" $\gamma = (\gamma_+,
\gamma_-)$ passing through slits and satisfying IP at~space-time points
where electron was measured.
Let us set
$$
\rho (\vec x_+, \vec x_-; t) = \!\!\!\!\!\!\!\!\sum_{\Sb
                                                 \gamma \in E\\
                                                 \gamma_\pm (1) = (t,
                                                 \vec x_\pm)
                                                 \endSb} \!\!\!\!\!\!\!\!
\Phi (\gamma_+, \gamma_-) = \sum_{k = 1}^n \psi_k (\vec x_+, t)
\psi_k^* (\vec x_-, t),
$$
where
$$
\psi_k (\vec x_+, t) = \!\!\!\!\!\!\!\!\sum_{\Sb
                                   \gamma \in A_k\\
                                   \gamma_+ (1) = (t, \vec x_+)
                                   \endSb}\!\!\!\!\!\!\!\!
\phi (\gamma_+).
$$
$\psi_k$ may be considered as a~(non-normalized) wave function
and~$\rho$ as the~(non-normalized) density matrix. Because
the~$\psi_k$'s are not normalized, we have means for constructing
general mixed states. For example, in~Exp.\,2, the~state of~an~electron
behind the~slits is described by $\sum_{k=1}^2 \psi_k (\vec x_+,
t) \psi_k^* (\vec x_-, t)$, where $\psi_k$ is the wave function
of~an~electron passing through the~$k$-th slit ($k = 1, 2$).

The interference character of~C-TP may be seen clearly from the
following property of~C-TP. In~C-TP, there are events with $\Phi(E) = 0$
having subevents $E_1 \subset E$ with $\Phi(E_1) \not= 0$ (in R-TP:
$P(E) = 0$, $E_1 \subset E \Rightarrow P(E_1) = 0$).
This is exactly what is observed in~QM: $\Phi(E(x))$ may be zero for
some $x \in \Bbb R$ but $\Phi (E_{kk}(x)) > 0$ and~$E_{kk}(x) \subset
E(x)$. Thus, the null-events $E$ in~C-TP are such that $\Phi(E_1) = 0$
for each $E_1 \subset E$.

$$
$$

\head
4. Theory of~Probability and Born's law
\endhead

Let us stress that the~Theory of~Probability $(\Omega, \Sigma, P)$ is
a~theoretical scheme, unverificable directly, especially the~quantity
$P(E)$ is not generally measurable, because this would be
the~measurenment of~the~probability of~an individual event and~this
is absurd. Only the relative frequency $p(E)$ of~an~"independently
repeated identical" event may be measured and~only for such an~event
does the~application of~the~Law of~large numbers (LLN) give the relation
$p(E) = P(E)$ [4].

The aim of~the Theory of~Probability (TP) can be formulated as follows:
to~find a~theoretical construction (using generally inverifiable
concepts) which enables us to~calculate the~relative frequencies
of~collective events. From this point of view a~certain similarity
between the~intuitive notion of~probability and~properties of~$P$ is
rather accidental and~irrelevant. This means that $P$, for example, need
not be a~real positive function, but the~relative frequency obtained
by~LLN must. We conjecture that both R-TP and~C-TP (including IP) form
a~priori possible bases of~a~description of~the~real world. The~quantum
mechanical experience shows undoubtedly that C-TP is a~true fundamental
TP. R-TP may be then considered as a~classical approximation to~C-TP
in~the~limit in~which the~interference effects (or non-real
C-probabilities) are negligible.

Let us now consider the~LLN in~C-TP. Let $E = E(x_0)$ be an~arrival
of~the~electron at~the point $x_0 \in \Bbb R$ on~a~screen. The
complementary event to~$E$ is $F = \bigcup_{x \not= x_0} E(x)$ and~$E, F
\in \Sigma_{mix}$.

The~C-probabilities $\Phi(E)$, $\Phi(F)$ are positive, but must be
normalized, because we are interested only in electrons measured
somewhere on~the screen; so let us set $\widetilde \Phi (E) = Z^{-1}
\Phi (E)$, $\widetilde \Phi (F) = Z^{-1} \Phi (F)$, $Z = \Phi (E \cup
F)$. Let $E_k$, $F_k$, $k = 1, 2, \dots$ be repetitions
of~the~events $E$ and~$F$. We shall assume that the~usual formula
$\Phi (E_k \cap F_l) = \Phi (E_k) \cdot \Phi (F_l)$ holds for independent
events. This formula holds for $\widetilde \Phi$ as well. Using
the~standard argument [4] from the~derivation of~LLN we obtain $p(E) =
\Phi (E)$. It follows from the~fact that $\Phi$ of~an~event $[|p(E) -
\Phi (E)| \geq \varepsilon > 0]$ tends to~zero and~that the~events with
$\Phi (E) = 0$ do not happen (this basic assumption is not usually
stated explicitly). In~terms of~usual QM (if~$E(x) \in \Sigma_{pure}$
for simplicity) we have
$$
\Phi (E(x)) = \Big| \sum_{\gamma_+(1) = x} \phi (\gamma_+) \Big|^2 =
\left| \psi(x) \right|^2,
$$
where the~wave function $\psi$ is not normalized. Thus, the~Born's law
will hold for the~normalized wave function
$$
\big| \widetilde \psi (x_0) \big|^2 = \widetilde \Phi (E(x_0)) = p(E(x_0)).
$$
We think that the~relative frequency $p$ should be used
in~the~formulation of~Born's law instead of~$P$. It is interesting that
in~our C-probabilitistic approach, this law is not a~postulate but
a~consequence.

$$
$$

\head
5. Interpretation postulate -- a~discussion
\endhead

We shall show that it suffices to assume the~validity of~IP
for~macroscopic systems only. This follows from the~hereditary
property of~IP: if IP holds for a~system, then IP holds for its subsystems.

Let us consider Exp.\,2 from Chapter 1. We shall  suppose that
the~electron going in~the~forward/backward time-direction interacts only
with photons going in~the~same time-direction. This agrees with the~fact
that quantum laws are written in~amplitudes $\phi$. The~interference
event (iii) now looks as follows (Fig.\,3),
where the gauge 1/0 detects the~scattered/not scattered photon. Using
IP for the~system \{electron + photon\}, we see that the~event described
above is not observable. It follows that the~event described above is
not observable. It follows that the~event (iii) in~Exp.\,2 (considering
the~electron as the~system and a~photon as a~measuring apparatus) cannot
happen for~the~system \{electron\} and this is exactly the~assertion
of~IP for~the~case of~\{electron\}.

$$
$$

\head
6. Straightforward physical interpretation and~the~C-Brownian motion
\endhead

Another interpretation of~C-TP may be given which is mathematically
equivalent but physically deeper than the~description given above.

Let us suppose (in the case of~non-relativistic QM) that there are two
sorts of~electrons -- the~forward and~backward ones with respect
to~the~time-direction of~their evolution and~that the~electron
of~different sorts move independently. To~the~event when
a~forward/backward electron moves along the~path $\gamma$ there
corresponds the~complex probability $\phi(\gamma) = ampl.$ for $\gamma /
\phi^* (\gamma)$. The~Interpretation Postulate must be refomulated.

IP': To the observation of~an~electron at~a~space-time point $x$ there
corresponds the~following event $E_x$. $E_x$ is the~event, when there
was simultaneously the~forward electron moving along $\gamma_+$
and~the~backward one moving along $\gamma_-$ such that both $\gamma_+$
and~$\gamma_-$ pass through $x$. The~formula
$$
\phi (E_x) = \sum_{\gamma_+, \gamma_- \in E_x} \phi (\gamma_+) \phi^*
(\gamma_-)
$$
holds because of~independence of~the~forward and~backward electrons.

Thus, for example, the physical transition $x \rightarrow y$ is
interpreted as an~event of~a~transition $x \rightarrow y$ of~a~forward
electron together with the~transition $y \rightarrow x$ of~some~backward
electron.

This interpretation allows us to~consider the~Feynman amplitude $\phi
(\gamma)$ directly as the~complex probability neglecting the~fact that
this is a~complex number, because the~amplitude $\Sigma \phi (\gamma)$
always will be positive for~an~observable event defined by~IP'.

The motion of~a~quantum mechanical particle may be considered as
the~complex analogue of~a~Brownian motion (this idea has already been
suggested but in~the~context of~R-TP [5]-[11]). Let us proceed further
with this idea: the~quantum particle can be viewed as an~object
scattered by~some medium (in~the~sense of~C-TP). Let us define
the~subquantum particle as the~"particle" unscattered by this medium and
we expect it be a~"point-like complex object". The~subquantum particle
moves in~the~"complex" medium and~its motion is influenced by a~large
number of~independent random effects, each of~them being very small.
The~"complex" medium should be interpreted as a~subquantum vacuum. These
questions will be considered in~detail in~the~next paper; let us mention
that the~analogy between the~propagator of~the~Schr\"{o}dinger equation
and~the~propagator of~the~Brownian particle and~the~analogy between
the~Schr\"{o}dinger and~Fokker-Planck equation has already been hinted
in~[12], [13], but our concept allows us to~consider the~Schr\"{o}dinger
equation {\it as} the~Fokker-Planck equation for~the~"complex" Brownian
particle.

Let us discuss briefly possible consequences of~this approach in~order
to~find a~possible deterministic interpretation of~QM.
\roster
\item"(i)"
   The probability amplitude (or wave function) is interpreted as the
C-probability, quantum superposition as an~addition of~C-probabilities,
and~the~principle of~indeterminacy as an~analogue
of~the~indeterminacy of~Brownian motion.
\item"(ii)"
   The hidden parameters may be defined as parameters describing
the~subquantum vacuum. These parameters are principially unobservable
(the~positions and~velocities of~atoms of~gas are also unobservable)
and~this causes the~statistical character of~QM. Theorems
on~non-existence of~hidden parameters [3] do not work here, because
these parameters are described by complex statistics.
\item"(iii)"
   So-called quantum jumps appearing at~a~measurement can be
understood in~"complex" statistics purely statistically.
The~act of~measurement is the~act of~finding which C-probable
possibility was realized, analogously to~the~case when the~position
of~a~Brownian particle is measured.
\item"(iv)"
   It can be expected that the~system \{particle + vacuum\} is
deterministic and that the~indeterministic character of~QM arises from
our neglecting the parameters that describe the~subquantum va\-cuum. There
is also another indeterminacy, because we can observe only certain
events described in~IP'. The~motion of~a~subquantum particle is
"C-Brownian-deter\-mi\-nistic" in~one direction of~the~time (i.e.
the~description by~$\phi (\gamma_+)$). On~the~other hand, the~physical
transition incorporates both directions of~time (i.e. $\phi(\gamma_+)
\phi^*(\gamma_-)$) which are mixed together.
\endroster
$$
$$

\head
7. Conclusions
\endhead

It follows from Bell's inequality and~from the~experiments described
in~[3] that the~local hidden parameter theory cannot explain
the~observed facts, providing that this theory is based on~the~R-theory
of~probability. In~the~C-TP Bell's inequality does not hold (since it is
based on~positivity of~probability). Thus the~local hidden parameter
theory using C-TP instead of~R-TP may solve the~above dilemma.

On~the~other hand, the~C-TP brings a~deeper and~more direct
understanding of~the~phenomenon of~quantum interference. The~main goal
of~the~paper was to~generalize the~notion of~probability in~such a~way
to~include quantum mechanics in~a~natural and~direct manner.

$$
$$

\head
Acknowledgements
\endhead

Sincere thanks are to~my brother, Dr. V. Sou\v{c}ek (Charles University,
Prague), who proposed the~physical interpretation of~C-TP in~terms
of~forward and~backward electrons. The~autor thanks Niels Bohr Institute
for their hospitality during his stay in~Copenhagen. I also thank
Eva Murtinov\'a for her help in typing of~this manuscript.



\enddocument
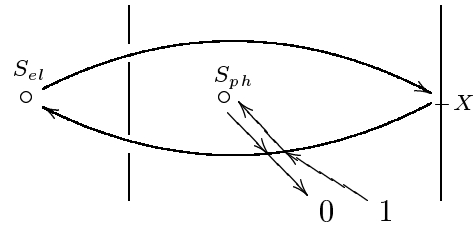
\bye